\documentclass[epj]{svjour}
\usepackage{graphicx}
\usepackage{amsmath}
\usepackage{amssymb}
\usepackage{mathtools}
\usepackage{xspace}
\usepackage{rotating}
\usepackage{multirow}
\usepackage{doi}
\usepackage{xspace}
\usepackage{xcolor}
\usepackage{float}
\usepackage{microtype}
\usepackage{siunitx}
\usepackage{floatflt}

\usepackage{hyperref}
\hypersetup{
	colorlinks=true,
    linkcolor=blue,
    filecolor=blue,
	urlcolor=blue,
    citecolor=blue,
}

\newcommand{\orcid}[1]{\href{https://orcid.org/#1}{\includegraphics[width=8pt]{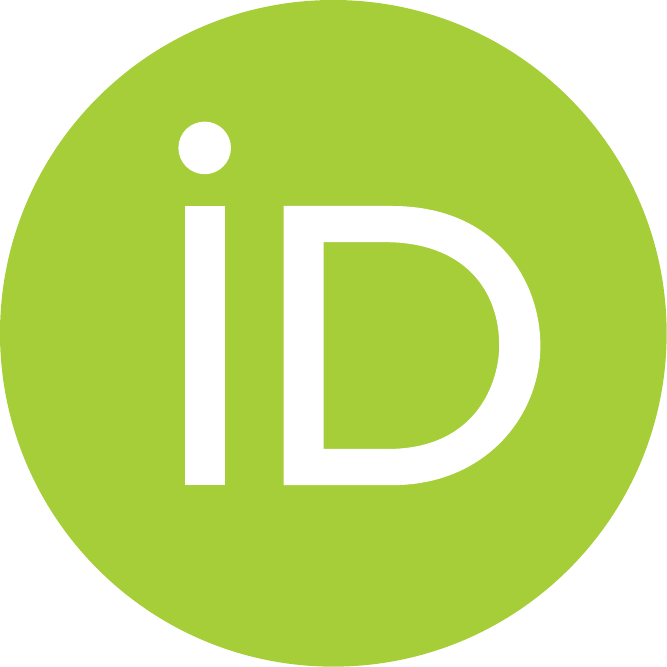}}}

\def\bbd{$2\nu\beta\beta$\xspace}
\def\onu{$0\nu\beta\beta$\xspace}

\begin{document}
	\pagestyle{plain}
%
\title{Array of Cryogenic Calorimeters to Evaluate the Spectral Shape of forbidden $\beta$-decays: the ACCESS project}
\author{L. Pagnanini\inst{1,2,3}\thanks{\emph{Corresponding Author:} lorenzo.pagnanini@gssi.it}\orcid{0000-0001-9498-5055}
\and G. Benato\inst{1,2}
\and P. Carniti\inst{4,5}
\and E. Celi\inst{1,2} 
\and D. Chiesa\inst{4,5}
\and J. Corbett\inst{3}
\and I. Dafinei\inst{1}
\and S. Di Domizio\inst{6}
\and P. Di Stefano\inst{3}\orcid{0000-0002-9709-6875}
\and S. Ghislandi\inst{1,2}\thanks{\emph{Corresponding Author:} stefano.ghislandi@gssi.it}\orcid{0000-0003-0232-1249}
\and C. Gotti\inst{5}\orcid{0000-0003-2501-9608}
\and D. L. Helis\inst{1,2}  
\and R. Knobel\inst{3} 
\and J. Kostensalo\inst{7}\orcid{0000-0001-9883-1256}
\and J. Kotila\inst{8,9,10}\orcid{0000-0001-9207-5824}
\and S. Nagorny\inst{3} 
\and G. Pessina\inst{5}
\and S. Pirro\inst{2}
\and S. Pozzi\inst{4,5}
\and A. Puiu\inst{2} 
\and S. Quitadamo\inst{1,2}\orcid{0000-0003-0107-1698} 
\and M. Sisti\inst{5}
\and J. Suhonen\inst{8}\orcid{0000-0002-9898-660X} (The ACCESS Collaboration)
\and S. Kuznetsov\inst{11}
} 

\institute{Gran Sasso Science Institute, L’Aquila I-67100, Italy
\and INFN – Laboratori Nazionali del Gran Sasso, Assergi (L’Aquila) I-67100, Italy
\and Department of Physics, Engineering Physics and Astronomy, Queen’s University, Kingston, ON, K7L 3N6, Canada
\and Dipartimento di Fisica, Universit\`a di Milano - Bicocca, Milano I-20126, Italy
\and INFN – Sezione di Milano Bicocca, Milano I-20126, Italy
\and Dipartimento di Fisica, Universit\`a di Genova and INFN Sezione di Genova, Genova I-16146, Italy
\and Natural Resources Institute Finland, Yliopistokatu 6B, FI-80100 Joensuu, Finland
\and Department of Physics, University of Jyv\"askyl\"a, P.O. Box 35, FI-40014, Jyv\"askyl\"a, Finland
\and Finnish Institute for Educational Research, University of Jyv\"askyl\"a, P.O. Box 35, FI-40014, Jyv\"askyl\"a, Finland
\and Center for Theoretical Physics, Sloane Physics Laboratory, Yale University, New Haven, CT 06520-8120, USA
\and Prokhorov General Physics Institute of the Russian Academy of Sciences
119991, Moscow, 38 Vavilov str., Russia}
\date{Received: date / Revised version: date}
%
\abstract{The \href{https://sites.google.com/gssi.it/access}{ACCESS} (Array of Cryogenic Calorimeters to Evaluate Spectral Shapes) project aims to establish a novel technique to perform precision measurements of forbidden $\beta$-decays, which can serve as an important benchmark for nuclear physics calculations and represent a significant background in astroparticle physics experiments. ACCESS will operate a pilot array of cryogenic calorimeters based on natural and doped crystals containing $\beta$-emitting radionuclides. In this way, natural (e.g. $^{113}$Cd and $^{115}$In) and synthetic isotopes (e.g. $^{99}$Tc) will be simultaneously measured with a common experimental technique. The array will also include further crystals optimised to disentangle the different  background sources, thus reducing the systematic uncertainty. In this paper, we give an overview of the ACCESS research program, discussing a detector design study and promising results of $^{115}$In.
}
\keywords{cryogenic calorimeters, beta decay, nuclear matrix element, neutrinoless double beta decay}
%
\maketitle
\section{Introduction}\label{Sec:Intro}
Since its debut in the scientific community, the neutrino has played a crucial role in the puzzle of fundamental particles, due to its peculiar cocktail of properties: at the same time the neutrino is the lightest massive particle of the Standard Model, the most elusive one, and the only neutral fundamental fermion. Particularly, this last feature makes the neutrino the perfect probe for Majorana's hypothesis, according to which a fermion and its own antiparticle have the same real wave function. Indeed, we can distinguish a neutrino from an antineutrino only indirectly, through its interactions in matter. They could be two different states of the same particle, which interacts in two different ways according to its chirality. The discovery of neutrino oscillations, revealing its massive nature, has pushed the experimental efforts in this direction, because only a massive particle can reverse its chirality. 

As far as we know, the only process that could unravel the mysterious nature of neutrino is the neutrinoless double beta decay (\onu). It is a forbidden, lepton-number-violating nuclear transition, whose observation would answer long-standing questions such as Dirac or Majorana nature of neutrinos and their absolute mass, also providing a possible leptogenesis mechanism to explain matter-antimatter asymmetry in the Universe.
The \onu would be an extremely rare decay, making its searches exceptionally challenging. In the case of light Majorana mass exchange mechanism, according to Fermi's golden rule, its half-life can be factorised as:
\begin{equation}\label{halflife}
\left( T_{1/2}^{0\nu} \right)^{-1} = g_\text{A}^4  \mathcal{G}^{0\nu}(Q_{\beta\beta}, Z)  \left\vert \mathcal{M}^{0\nu}(A,Z) \right\vert^2  \left\vert \frac{m_{\beta\beta}}{m_e} \right\vert^2
\end{equation}
where $\mathcal{G}^{0\nu}(Q_{\beta\beta}, Z)$ is the phase-space factor, $g_{\rm A}$ is the axial-vector coupling strength, $\mathcal{M}^{0\nu}(A,Z)$ is the nuclear matrix element (NME), $m_e$ is the electron mass, and $m_{\beta\beta}$ is the effective Majorana mass of the neutrino. The latter is the New Physics parameter that we can derive from Eq.~(\ref{halflife}), provided that we know all the involved variables. The phase-space factor depends on the decay kinematics, and its value is calculated with a high precision for each isotope~\cite{Kotila:2012zza,Stoica:2013lka}. On the contrary, the uncertainties of the theoretical estimates for the NME are large, at least of the order of 30\% based on the variation between the NME values provided by various theoretical frameworks adopting different approaches to solve the nuclear many-body problem~\cite{Engel_2017,EjiriNME,Barea:2015kwa}.

Further evidence that nuclear theory has room for improvement is the fact that a {\it quenching} of the free-nucleon axial coupling constant, $g_{\rm A}^{\rm free} = 1.27641$~\cite{gAvalue}, is usually required to bring theory in agreement with $\beta$-decay experimental data within some theoretical models. The same holds for two-neutrino double beta decay (\bbd)~\cite{Barea:2015kwa}, the standard-model-allowed counterpart of the \onu. Whether the quenching found for $\beta$-decay and \bbd applies similarly to the \onu is still a matter of debate~\cite{Engel:2016xgb,EJIRI20191}; indeed, these nuclear processes have a low-momentum exchange while \onu is a high-momentum exchange process involving states up to $\sim$100 MeV. Though it is not obvious how the quenching in \bbd generalises to \onu, it is of utmost importance to study the quenching of $g_{\rm A}$ with different approaches to identify its origin and to understand the implications of it when probing new physics phenomena.

In this scenario, new experimental data to benchmark nuclear model calculations are pivotal. Indeed, if a next-generation experiment observes the \onu decay, the NME uncertainty will directly reflect on the Majorana mass, and on the half-lives predicted for other isotopes. On the other hand, if next-generation experiments fail to observe \onu decay, the scientific community would know which are the most favoured isotopes to steer the subsequent efforts, but again NME uncertainties complicate the selection process. For this reason, the Astro Particle Physics European Committee (APPEC) in its recent report on European Strategy for \onu recommends \textit{``A dedicated theoretical and experimental effort, in collaboration with the nuclear physics community to achieve a more accurate determination of the NMEs''}~\cite{Giuliani:2019uno}. This is exactly what the ACCESS project aims to do.

There are several ways to quantify the effective quenching in decay processes with low-momentum exchange such as discussed in Refs.~\cite{Engel:2016xgb,Suhonen:2017krv}. One of the proposed methods exploits the dependence of the $\beta$-spectrum shape of forbidden non-unique beta transitions on the $g_{\rm A}$ value. Since in these particular decays high multipolarities are involved, as in \onu, they are an attractive tool to investigate the nuclear aspects of this transition. A detailed review of theoretical and experimental achievements in this field can be found in Ref.~\cite{EJIRI20191}. Moreover, several experimental searches in astroparticle physics are affected by assumptions regarding the $\beta$-decay spectral shape, such as the $^{210}$Bi in solar neutrino detection~\cite{BOREXINO:2020aww}, $^{39}$Ar in dark-matter search with liquid argon-based detectors~\cite{DarkSide-50:2022qzh}, and $^{90}$Sr in cryogenic calorimeters~\cite{Azzolini:2019yib,Armengaud:2019rll,Adams:2020dyu}. A precise evaluation of the spectral shape of these background components would notably reduce the systematic uncertainty in the data modelling and improve the experimental sensitivity.

In this field, the main experimental efforts focused on naturally occurring isotopes for which a forbidden non-unique $\beta$-decay is expected: $^{113}$Cd, $^{115}$In, and $^{50}$V. $^{113}$Cd $\beta$-decay is the most studied process, thanks to well-established experimental methods. The most accurate measurement was performed by the KINR-DAMA collaboration at the Laboratori Nazionali del Gran Sasso (LNGS) by using a CdWO$_4$ scintillating crystal in long-term low-background measurements. The shape of the spectrum was precisely measured with an energy threshold of 28 keV, an energy resolution of 47 keV at $Q_{\beta} = (323.83 \pm 0.27)$ keV~\cite{wang2017ame2016}, and a signal-to-background ratio of $\sim$56~\cite{Belli:2007zza}. Recently, the COBRA collaboration carried out a very precise measurement with CdZnTe (CZT) semiconductor detectors, achieving an energy threshold of 84 keV, an energy resolution of 18 keV at $Q_{\beta}$, and a signal-to-background ratio of $\sim$47~\cite{Bodenstein-Dresler:2018dwb}.
$^{50}$V presents an opposite situation, having its $\beta$-decay not yet observed. Different experimental approaches are under development~\cite{Laubenstein:2018euc,Pattavina:2018nhk}
to investigate such decay beyond the current limit, i.e., $T_{1/2} > 1.9 \times 10^{19}$ yr~\cite{Laubenstein:2018euc}.
For $^{115}$In, there were two measurements based on scintillating liquid loaded with In~\cite{PhysRev.122.1576,PhysRevC.19.1035}, and LiInSe$_{2}$ cryogenic calorimeter~\cite{Leder:2022beq}. While the former measurement featured a poor energy resolution and energy threshold which prevented a detailed study of the spectral shape, this was successfully performed in the latter case. This is a further confirmation that cryogenic macro-calorimeters are a proper tool to asses the spectral shape of $\beta$-decay, as extensively demonstrated in last years for the two-neutrino double beta decay~\cite{Azzolini:2019yib,Armengaud:2019rll,Adams:2020dyu,CUPID:2019kto}. The success of these efforts led to the idea of applying cryogenic calorimeters to spectral shape measurements of rare $\beta$-decays. There are other alternative or complementary experimental approaches based on Metallic Magnetic Calorimeters~\cite{LOIDL2019108830} or Silicon Drift Detectors~\cite{Gugiatti:2020wad,NAVA2023167812} which are consolidating, reflecting the growing interest in this research field.

In this work, we introduce the ACCESS project, providing a general overview of the research program. In Sec.~\ref{Sec:Theory} we introduce the spectral shape method, and in Secs.~\ref{Sec:Xtals} and \ref{Sec:Sensors} we detail the features of crystals and sensors, respectively. In Sec.~\ref{Sec:MC} we describe the Monte Carlo techniques adopted for the ACCESS purpose, in particular for the performance study presented in Sec.~\ref{Sec:Study}. Finally, we discuss the preliminary results of the ACCESS measurements in Sec.~\ref{Sec:Measurements}, and the project perspectives in Sec.~\ref{Sec:Conclusions}.

\section{Theoretical framework}\label{Sec:Theory}
The Spectral Shape Method (SSM)~\cite{Haaranen:2016rzs} is based on the fact that for forbidden non-unique $\beta$-decays the energy spectrum of the emitted electrons -- described by the $\beta$-decay shape function -- depends on the leptonic phase-space factors and the Nuclear Matrix Elements (NMEs) in a non-trivial way.
The phase-space factors can be calculated to a desired accuracy~\cite{Mustonen:2006qn}, but the NMEs are subject to systematic uncertainties of the underlying nuclear models.

The complexities of the shape function are condensed in the so-called shape factor $C(w_e)$, which is a function of the total energy $w_e$ of the emitted electron (in units of the electron rest mass)~\cite{Suhonen:2017krv,Haaranen:2016rzs}.
The shape factor can be divided into an axial-vector part $C_{\rm A}(w_e)$, a vector part $C_{\rm V}(w_e)$, and a mixed vector-axial-vector part $C_{\rm VA}(w_e)$, such that we obtain
\begin{equation}
C(w_e) = g_{\rm A}^2\Big[C_{\rm A}(w_e) + \left(\frac{g_{\rm V}}{g_{\rm A}}\right)^2C_{\rm V}(w_e) + \frac{g_{\rm V}}{g_{\rm A}}C_{\rm VA}(w_e) \Big] \,.
\label{eqn:shape_factor}
\end{equation}
Here the interference between the sum of the $C_{\rm A}$ and $C_{\rm V}$ parts and the mixed $C_{\rm VA}$ part causes a variation of the electron spectral shapes depending on the ratio $g_{\rm V}/g_{\rm A}$.

In~\cite{Haaranen:2016rzs} it was noticed that the variation was rather strong for the $\beta$-decay transition $^{113}\textrm{Cd}(1/2^+) \to \,^{113}\textrm{In}(9/2^+)$ when the NMEs were computed by using the interacting shell model (ISM)~\cite{RevModPhys.77.427} and the microscopic quasiparticle-phonon model (MQPM)~\cite{Toivanen:1998zz}.
In a further study of this transition~\cite{Haaranen:2017ovc}, a similar dependence was recorded for the microscopic interacting boson-fermion model (IBFM-2)~\cite{iachello_isacker_1991}.
The essence of the SSM philosophy is to compare the electron spectral shapes, computed for different values of $g_{\rm A}$, with the measured one in order to be able to access an effective value of $g_{\rm A}$. In Ref.~\cite{Haaranen:2017ovc} the computed electron spectra are compared with the one measured in Ref.~\cite{Belli:2007zza}, finding that the measured spectrum is roughly reproduced by the calculated spectra of all three nuclear models for values of $g_{\rm A}/g_{\rm V}\sim 0.9$.
This is a remarkable result if we consider the totally different nuclear-structure principles behind these models.

In the studies~\cite{Haaranen:2016rzs,Haaranen:2017ovc}, as also in Ref.~\cite{Bodenstein-Dresler:2018dwb}, it was noticed that a simultaneous prediction of the electron spectral shape and the partial half-life of the decay transition was a challenge. In Ref.~\cite{Kumar2020}, it was proposed to use the conserved-vector-current hypothesis (CVC) related NME, introducing the so-called small relativistic NME~\cite{Behrens1982} (here denoted as ``s-NME"), to fix the problem with the prediction of the partial half-life of the decay transition simultaneously with the $\beta$-decay spectral shape. This ``enhanced SSM" was successfully tested in the work~\cite{Kostensalo2021} for the $^{113}$Cd decay using the three mentioned nuclear models, and further applications were presented in~\cite{Kumar2021}. To the recently measured $^{115}$In decay~\cite{Leder:2022beq} the enhanced SSM was not yet applied, but in the ACCESS project we are going to harness the full power of this method in order to obtain fine-tuned results for $^{115}$In and other potentially interesting nuclei.

The value of the s-NME is directly related to the CVC in an ideal calculation ~\cite{Behrens1982}. The magnitude of the CVC-based s-NME is typically 2-3 orders of magnitude smaller than the magnitude of a typical main NME which collects strong contributions from the $1\hbar\omega$ shell-model valence space~\cite{Kumar2021}. Instead, the s-NME collects its contributions from beyond the $1\hbar\omega$ valence space, making the prediction of its value very hard for the nuclear models used in the SSM calculations. In particular, due to their very restricted valence spaces, the ISM and IBFM-2 typically predict a zero value for the s-NME~\cite{Kumar2021}. The MQPM, instead, has a large valence space but still it predicts too small magnitudes for the s-NME. Despite its small value, the s-NME plays a role in determining the half-life of a beta transition. This is why it is important to have an estimate of its value, and its CVC value is a good reference in the enhanced SSM calculations.

Since the CVC value of the s-NME requires an ``absolute" calculation (infinite valence space, perfect many-body theory, etc.) it cannot be used directly in realistic applications. Instead, it can be used as a free parameter and its value can be adjusted to reproduce the experimental partial half-life of the beta transition of interest. This adjustment has to be done for each value of $g_{\rm A}$ separately so that the computed beta-spectrum templates used in the enhanced SSM will fix both $g_{\rm A}$ and s-NME when matched with the experimental beta spectrum. The thus fixed value of the s-NME is typically in the ballpark of its CVC value~\cite{Kumar2021}.

\section{Crystals}\label{Sec:Xtals}
The ACCESS project proposes to use different crystals to measure the spectral shape of different $\beta$-emitters. A list of isotopes interesting for Nuclear and Astroparticle Physics is presented in Tab.~\ref{tab:isotopes}.

\begin{table}[ht]
\centering
\caption{List of the isotopes whose $\beta$-decay could be measured using the carrier crystal approach proposed by ACCESS (in bold) or natural crystals. In the rightmost column, we report the isotopic abundance of naturally occurring isotopes, and the target activity of artificial isotopes in doped crystals. $^{210}$Pb and $^{210}$Bi belong to $^{238}$U natural radioactive chain, so that their spectra can be measured with natural PbWO$_{4}$ exploiting the residual $^{210}$Pb contamination}

\begin{tabular}{ccccc}
\hline
\hline
\textbf{Physics Case} & \textbf{Isotope} & \textbf{Q$_{\beta}$} & \textbf{Half-life} & \textbf{Natural Abundance}\\
                                 &       & [keV] & [yr] & \textbf{or Target Doping}\\
                                 \hline
\multirow{3}{*}{Nuclear Physics} & \textbf{$^{99}$Tc} & 293.8 & $2.11 \times 10^{5}$ & 0.25 ppb\\
                                 & $^{113}$Cd & 316 & $7.70 \times 10^{15}$ & 13.47 \%\\
                                 & $^{115}$In & 496 & $4.41 \times 10^{14}$ & 95.7 \%\\
                                 \hline
                                 & \textbf{$^{90}$Sr} & 545.9 & 28.8 & 30 ppq \\
Background in $\nu$-physics      & \textbf{$^{39}$Ar} & 565  & 269 & 0.15 ppt\\
and Dark Matter search           & \textbf{$^{42}$Ar} & 599  & 32.9 & 20 ppq\\
                                 & $^{210}$Bi & 1161.2 & 0.014 & $^{238}$U decay chain\\
                                 \hline
Cosmic Neutrino background       & \textbf{$^{151}$Sm} & 76.4 & 94.7 & 0.20 ppt\\
detection                        & $^{210}$Pb & 63.5 & 22.2 & $^{238}$U decay chain\\
    \hline
    \hline
    \end{tabular}
    \label{tab:isotopes}
\end{table}

{\bf Natural crystals.} Only three forbidden $\beta$-decay emitters are natural isotopes with a high isotopic abundance (i.a.); $^{115}$In  (i.a. $= 95.72\%$), $^{113}$Cd (i.a. $= 12.23 \%$), and $^{87}$Rb (i.a. $= 27.835 \%$). In particular, the first two isotopes can be embedded in crystals suitable to be used as an absorber in a cryogenic calorimeter, such as indium oxide (In$_2$O$_3$)~\cite{Celi:2021dhe}, indium iodide (InI), and cadmium tungstate (CdWO$_4$). These are the first crystals and isotopes to which ACCESS will focus on. Another interesting candidate to explore as cryogenic calorimeter for the ACCESS purposes could be the PbWO$_4$, the material used for the electromagnetic calorimeter of the Compact Muon Solenoid (CMS)~\cite{CMS_ECAL} experiment at Large Hadron Collider (LHC). The residual amount of $^{210}$Pb from natural radioactivity in such crystal could allow one to study the low-energy $\beta$-decay of $^{210}$Pb, and the subsequent decay of $^{210}$Bi. This material is currently under investigation for RES NOVA, a proposed experiment to detect Supernova Neutrinos with PbWO$_4$ based cryogenic calorimeters at LNGS~\cite{RES-NOVA:2021gqp,RES-NOVAGroupofInterest:2022glt}.

{\bf Doped crystals.} Unfortunately, natural crystals would allow to study only a limited number of isotopes belonging to the long list of interesting candidates. A different approach is therefore needed to develop a technique able to assess simultaneously a higher number of isotopes. Based on previous successful experience with doped crystal to study $\alpha$-decay~\cite{Casali:2016vbw}, we propose to have a {\it carrier crystal}, e.g. TeO$_2$ or Li$_2$MoO$_4$, doped with a specific $\beta$-decaying isotope during the crystal growth. For each isotope, a twin pair of crystals, one doped and one natural can be operated simultaneously. While the embedded $\beta$-source will dominate the counting rate of the doped crystal, the natural one will monitor internal and environmental backgrounds. A combined fit of Monte Carlo simulations to the two spectra will ensure a precise reconstruction of both signal and background spectral shapes. For each isotope, the crystal size will be determined by balancing the containment efficiency of the $\beta$-signal, and the background rate in the region of interest. Given the Q-value of the considered processes ($< 600$ keV), a 1 cm$^3$ TeO$_{2}$ crystal would be a reasonable candidate as a carrier crystal. Therefore, we use this as a reference for our $\beta$-source concentration calculations. Indeed, the powder doping can be tuned in order to optimise the signal rate (e.g. $\sim2$ Hz) without spoiling the bolometric performance. As reported in Tab.~\ref{tab:isotopes}, the target doping to get a signal rate of $\sim2$ Hz in a 1 cm$^3$ TeO$_{2}$ crystal is at most at ppb level, giving a negligible contribution to the detector heat capacity. The main challenge of this approach is to obtain uniform doping in the crystal volume to ensure a correct evaluation of the detection efficiency. For this reason, several doping tests will be needed to study source segregation, and eventually, the optimal configuration will be reached after several growths. The goal of ACCESS is to demonstrate the potential of this technique at least with an isotope, reasonably $^{99}$Tc.

\section{Thermal sensors}\label{Sec:Sensors}
\begin{figure}[ht]
	\centering
	\includegraphics[width=0.85\textwidth]{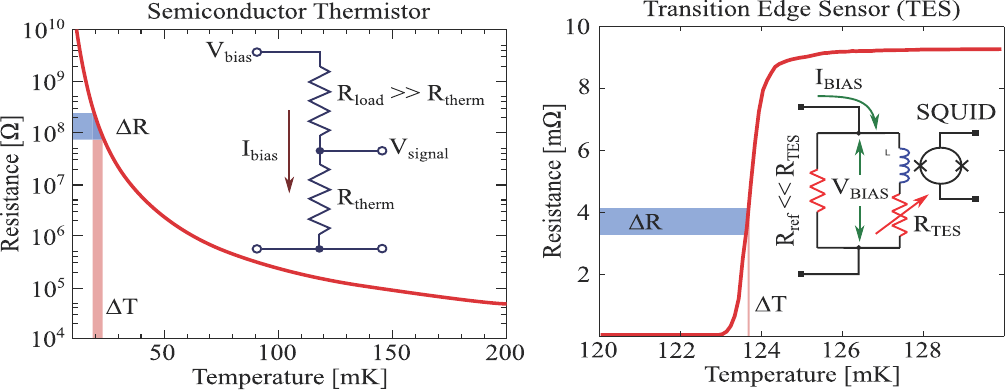}
	\caption{Resistance as a function of the NTD (left) and TES (right) temperature. A simplified scheme of the bias circuit for the two sensors is also shown. $R_\text{therm}$ and $R_\text{TES}$ represent the resistance of the NTD and the TES, respectively. Figure adapted from~\cite{Giachero_2017}}
	\label{fig:NTD_TES}
\end{figure}

The natural choice for the ACCESS thermal sensors is semiconductor thermistors, i.e. Neutron Transmutation Doped germanium (Ge-NTD or NTD)~\cite{Haller1984}, developed and successfully operated by CUORE, CUPID-0 and CUPID-Mo collaborations. Ge-NTDs are obtained by doping an ultra-pure germanium slab up to a concentration close to the critical value for the metal-insulator transition~\cite{Mott}. In this condition, below $\simeq$ 1 K, the electrical conductivity is driven by the Variable-Range Hopping (VRH) mechanism~\cite{doi:10.1080/14786436908216338}, and the relation between electrical resistivity ($\rho$) and temperature ($T$) is described by the Mott law~\cite{VRH_Elfros}: 
\begin{equation}
	\rho (T) =  \rho_0 \exp \left(\dfrac{T_0}{T} \right)^{\frac{1}{2}}
		\label{eqn:NTD_resistivity}
\end{equation}
$\rho_0$ is a parameter that depends on the intrinsic properties of the germanium lattice, while $T_0$ depends on the uniformity and concentration of the doping. Through the variation of its resistance (Fig.~\ref{fig:NTD_TES} left), the NTD converts a temperature variation, induced by an energy deposition in the crystal, into an electric signal.

The NTD is operated by biasing it with a potential difference $V_{bias}$ through an electric circuit where each sensor is connected in series with a load resistor $R_{load} \gg R$, as reported in Fig.~\ref{fig:NTD_TES} (left). In this way, for any bias voltage $V_{bias}$, the current flowing through the thermistor can be assumed to be constant. The optimal working point is chosen maximising the SNR by scanning a voltage region where the NTD response is still linear.

In underground astroparticle physics, NTD-based cryogenic calorimeters are usually operated at temperatures of the order of \SI{10}{mK} to reduce the heat capacity of the absorber, thus enhancing the signal amplitude. NTDs feature a high dynamic range, allowing them to detect energy depositions over a wide interval, from few keV up to $\simeq$ 10 MeV, and a very good energy resolution, of the order $\simeq 0.2-0.3 \%$~\cite{CUORE:2021mvw,Augier:2022znx}. However, they are characterized by a relatively slow time response, so thermal pulses develop over a time interval that ranges from hundreds of ms up to few seconds (depending also on the heat capacity of the crystals)~\cite{CUORE:2021mvw}. Therefore, a bolometer read out by NTDs can be affected by pile-up between different thermal pulses. Since the energy of two piled-up signals cannot be precisely reconstructed, these events are rejected reducing the selection efficiency in high-rate measurements ($>$ 1 Hz)~\cite{Leder:2022beq}.

In natural crystals, the signal counting rate is determined by the crystal dimensions, which can be optimised to increase the signal-to-background ratio. Since the expected counting rate due to $\beta$-signal and radioactive backgrounds in this application is $\mathcal{O}$(\SI{200}{mHz}), NTD readout is very suitable. Indeed, with a time resolution at a level of few ms and a time acquisition window of 500 ms~\cite{Celi:2021dhe}, the pileup probability is below 10\%.

Doped crystals exhibit a signal counting rate dependent on the source concentration, which can be adjusted by balancing the high signal-to-background ratio and the low pileup probability. In this case,  NTD sensor may not be the most effective choice, because the slow time response limits the maximum tolerable activity to $\mathcal{O}$(1 Hz). The Transition Edge Sensors (TES) can overcome such limitations. 

A TES is a thin film made of a superconducting material, operated at a temperature just below its superconducting transition temperature so that its resistivity is negligible. When an energy deposition occurs in the absorber to which the TES is coupled, the induced rise in temperature causes the resistance of the TES to increase, following the superconductive-to-normal metal transition (see Fig.~\ref{fig:NTD_TES} right). Since the transition lies in a temperature range of $\sim$ \SI{1}{mK}, the TES is extremely sensitive also to very small energy depositions, featuring excellent energy resolution and extremely low energy threshold. The values of these parameters depend a lot on the different sensor optimisations performed for the specific physics case~\cite{Irwin2005,Ullom_2015,Koehler:2021fqj,Singh:2022rck,Angloher:2016hbv,STRAUSS2017414,COSINUS:2021onk}. 
Another advantage of TES is that they are sensitive to athermal phonons rather than thermal phonons (like is the case for NTDs), therefore their time response is faster than the one of the NTD, but also, in this case, the actual value depends on the specific application. 
For the purposes of this paper, we assume the TES performance obtained in Refs.~\cite{STRAUSS2017414,COSINUS:2021onk} and reported in Tab.\ref{tab:detectorPerformance}, because they have been achieved with detectors similar to the ACCESS ones.
For ACCESS purposes, it would be very useful to combine the two temperature sensors operating them together and in two different dynamic ranges. The NTD readout provides data useful for the internal background reconstruction via $\alpha$-spectroscopy while data collected with TES readout allows mitigating the pileup in the signal.
Operating NTD and TES on the same crystal featuring a high counting rate would also enable the possibility to study and improve the pileup rejection efficiency achieved with NTD, which is of pivotal importance for the CUPID \onu next-generation experiment~\cite{CUPID:2020cpe,CUPID:2020smz}.
\begin{table}[ht]
\centering
\caption{Detector performance used in the Monte Carlo simulation processing for NTD~\cite{Nagorny:2019syb} and TES sensors~\cite{COSINUS:2021onk,STRAUSS2017414}. These parameters can vary from one detector to another depending on the features of sensor, absorber, and their coupling. For this reason, we chose as reference performance that of detectors as similar as possible to those foreseen for ACCESS}
\begin{tabular}{rll}
\hline
\hline
\textbf{Parameter} & \textbf{NTD} & \textbf{TES} \\
\hline
Rise Time & 10 ms & 1 ms\\
Acquisition window & 1 s & 100 ms\\
Energy resolution & 1 keV & 100 eV\\
Energy threshold & 5 keV & 1 keV\\
    \hline
    \hline
    \end{tabular}
    \label{tab:detectorPerformance}
\end{table}

\section{Monte Carlo simulations}\label{Sec:MC}
Given a certain $\beta$-decaying isotope to be measured, the corresponding detector design can be optimised by means of Monte Carlo simulations. Hereafter, we describe the software tools we plan to use for this purpose, while their application to the sample case study of $^{115}$In is reported in Sec.~\ref{Sec:Study}.

We simulate the signal and the different background sources with a Monte Carlo toolkit, called \textit{Arby}, based on the \textsc{Geant4} framework~\cite{Geant4}, version 4.10.03. 
The radioactive decay from the various sources can be generated in any volume or surface of the detector, cryostat and shielding implemented in \textit{Arby}~\cite{Azzolini:2019nmi}.
The primary and any secondary particles are then propagated through the detector geometry using the Livermore physics list. 
The energy deposited in crystals is recorded in the Monte Carlo output together with the time at which the interaction occurred. 
The fraction of energy released by any particle type is also recorded to allow particle identification.
Radioactive decays are implemented using the G4RadioactiveDecay database. The decay chains of $^{232}$Th, $^{238}$U, and $^{235}$U can be simulated completely or in part, to reproduce breaks of secular equilibrium. The $\beta$-decay simulations are generated according to the energy spectral shape templates, obtained as described in Sec.~\ref{Sec:Theory}.

In order to implement the detector response function and data production features in the Monte Carlo data, we reprocess the \textit{Arby} output with a dedicated code. In particular, to account for detector time resolution, we sum energy depositions that occur in the same crystal within the signal rise time. We reproduce the single signal selection defining a pile-up window\footnote{The pile-up window reproduces the experimental data time acquisition window, which is not always symmetrical with respect to the trigger position (see Fig.~\ref{fig:Spectrum+Pulse} left).} around each event: if another signal falls in it we discard both the events as in the data processing. The experimental energy resolution is usually reproduced by applying a Gaussian smearing function with linearly variable width based on measured FWHM of $\gamma$-ray and $\alpha$-particle peaks, while the energy threshold of each detector is modelled with an error function that interpolates the experimental data of trigger efficiency versus energy. In the specific case of the study presented in Sec.~\ref{Sec:Study}, we assume a constant energy resolution ($\sigma$) and a conservative energy threshold of 5$\sigma$. The selection and detection efficiencies are calculated on the data as a function of the energy, and modelled with an error function which is applied to the Monte Carlo simulations.

\section{Detector design studies: the In-115 case} \label{Sec:Study}
With the Monte Carlo tools just described, we conduct a study to optimise the detector design of an indium iodide crystal. 
We evaluate the side size of the crystal by optimising the interplay between signal and background rates.
The signal simulation takes as input a theoretical energy spectrum template with fixed $\text{s-NME}=2.0$ and $g_{\rm A}=0.9$, which are assumed to be reasonable values according to a previous enhanced-SSM analysis on $^{113}$Cd~\cite{Kostensalo2021}. For this study, we consider in our simulations InI cubic crystals of different sides in the range 2 - 10 mm, computing the expected activity accordingly:
\begin{equation}
A(Bq) = \frac{N_0}{\tau} = \frac{d\cdot l^3 \cdot N_A \cdot i.a.}{M\cdot\tau}
\end{equation}
where $d$ is the InI density, $l$ the crystal side, $N_{A}$ is the Avogadro number, $M$ the InI molar mass, while $i.a.$ and $\tau$ are the isotopic abundance and the mean-life of $^{115}$In, respectively.

On the background side, we expect $^{232}$Th, $^{238}$U chains and $^{40}$K to be the most prominent contributions. Therefore, we simulate these radioactive sources in the thermal shields of the cryostat, characterised by the CUORE-0 and CUPID-0 experiments~\cite{Azzolini:2019nmi}, and in the crystal bulk. For the latter, we compute quite conservative limits exploiting the InI data shown in Sec.~\ref{Sec:Measurements}.
Typical performance of detectors similar to the ones we expect to use are reported in Tab.~\ref{tab:detectorPerformance}.

\begin{figure}[ht]
	\centering
	\includegraphics[width=0.48\textwidth]{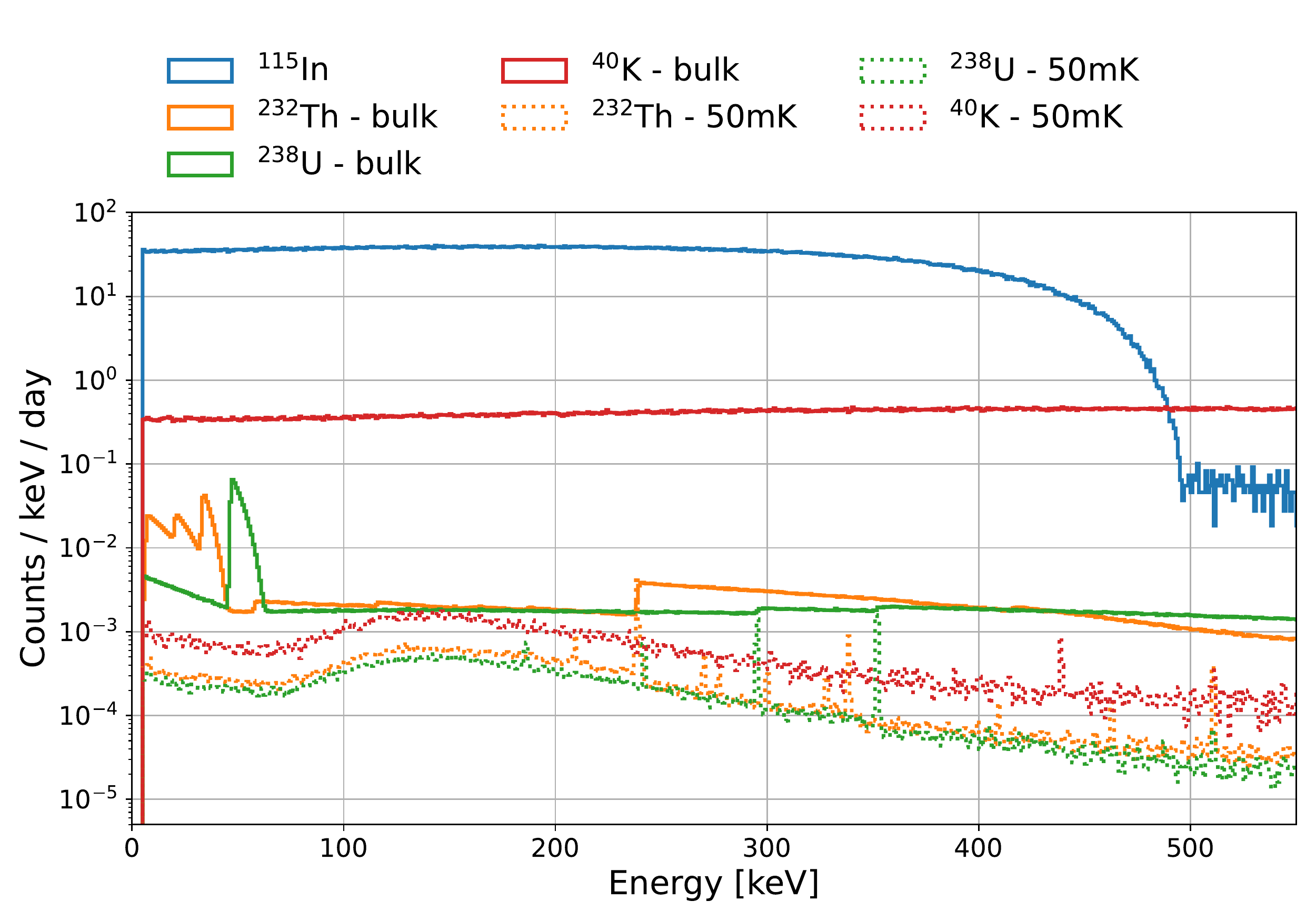}
	\caption{Simulated energy spectra of the $^{115}$In $\beta$-decay (blue), and the different background components for a \SI{7}{mm}-side InI crystal with NTD readout. We include in this plot only the dominant background contributions, i.e. the bulk contaminations of the crystal and the thermal shield at 50 mK. As expected, the $^{115}$In $\beta$-decay is two orders of magnitude higher with respect to the limit on the $^{40}$K background (solid red)}
	\label{fig:totalBM}
\end{figure} 

\begin{figure}[ht]
	\centering
	\includegraphics[width=0.48\textwidth]{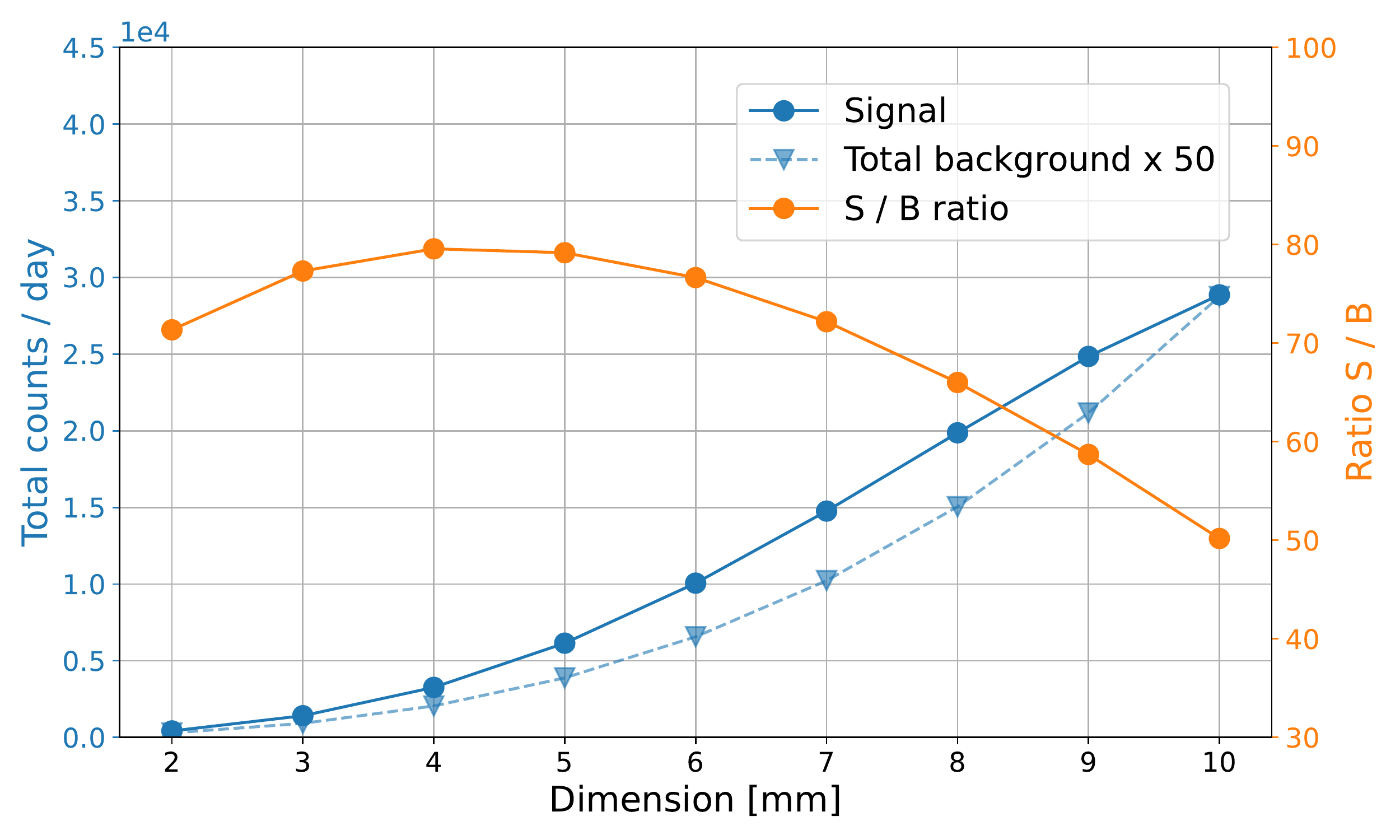}
	\caption{Signal rate (blue solid line), limit on background rate (blue dashed line), and signal-to-background ratio (orange solid line) as a function of the absorber side. The signal rate grows with the dimension even if, for larger sizes, the occurrence of multiple decay within the same acquisition time window reduces the efficiency of signal selection. The limit on the background increases almost linearly with the crystal volume, thus the signal-to-background ratio is more favourable for small crystals}
	\label{fig:SBRatio}
\end{figure}

The resulting energy spectra for an InI crystal of size \si{7}$\times$\si{7}$\times$\si{7} mm$^{3}$, shown in Fig.~\ref{fig:totalBM}, demonstrates that the expected background level is low enough to ensure a clear assessment of the $\beta$-decay spectral shape. In order to understand which are the optimal crystal dimensions, for each configuration, we integrate the energy spectrum of signal and background in the energy interval from \SI{0}{keV} up to the Q-value of the $^{115}$In $\beta$-decay. The resulting signal and background throughput, together with the signal-to-background ratio, are reported in Fig.~\ref{fig:SBRatio}. 

The signal rate increases with the size of the crystal due to a larger indium activity, allowing the collection of more statistics. However, the effect of the pile-up is not negligible for larger absorbers, thus damping the previous trend. 
This happens when the mean time between subsequent events becomes comparable with the acquisition window length of \SI{1}{s}.
The background rate, instead, being dominated by bulk contamination, always gets larger with the crystal dimensions. 
As a result, the signal-to-background ratio reaches its maximum at absorber size, of $\sim$ \SI{4}{mm} and then decreases, as shown in Fig.~\ref{fig:SBRatio}. 
Thus, little absorbers are preferred from the signal-to-background point of view. 
However, larger absorbers, beside the higher absolute signal rate, are characterised by a higher containment efficiency, useful in the fit procedure to better identify and fix background peaks.
Taking everything into account, we consider a crystal of \SI{7}{mm} side size a good compromise to maximise the signal rate and the background modelling efficacy while keeping a satisfying signal-to-background ratio.

\begin{figure}[ht]
	\centering
	\includegraphics[width=0.48\textwidth]{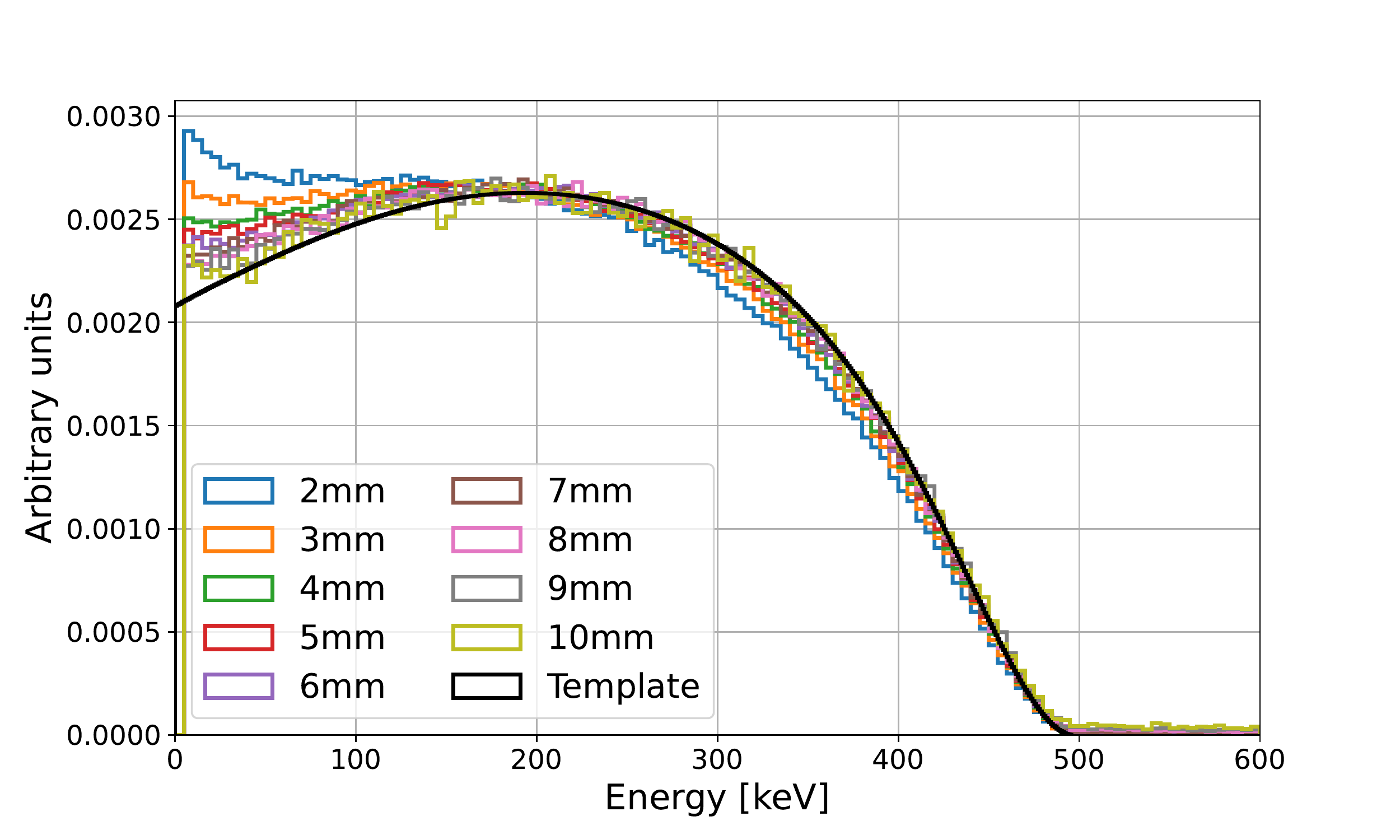}
	\includegraphics[width=0.48\textwidth]{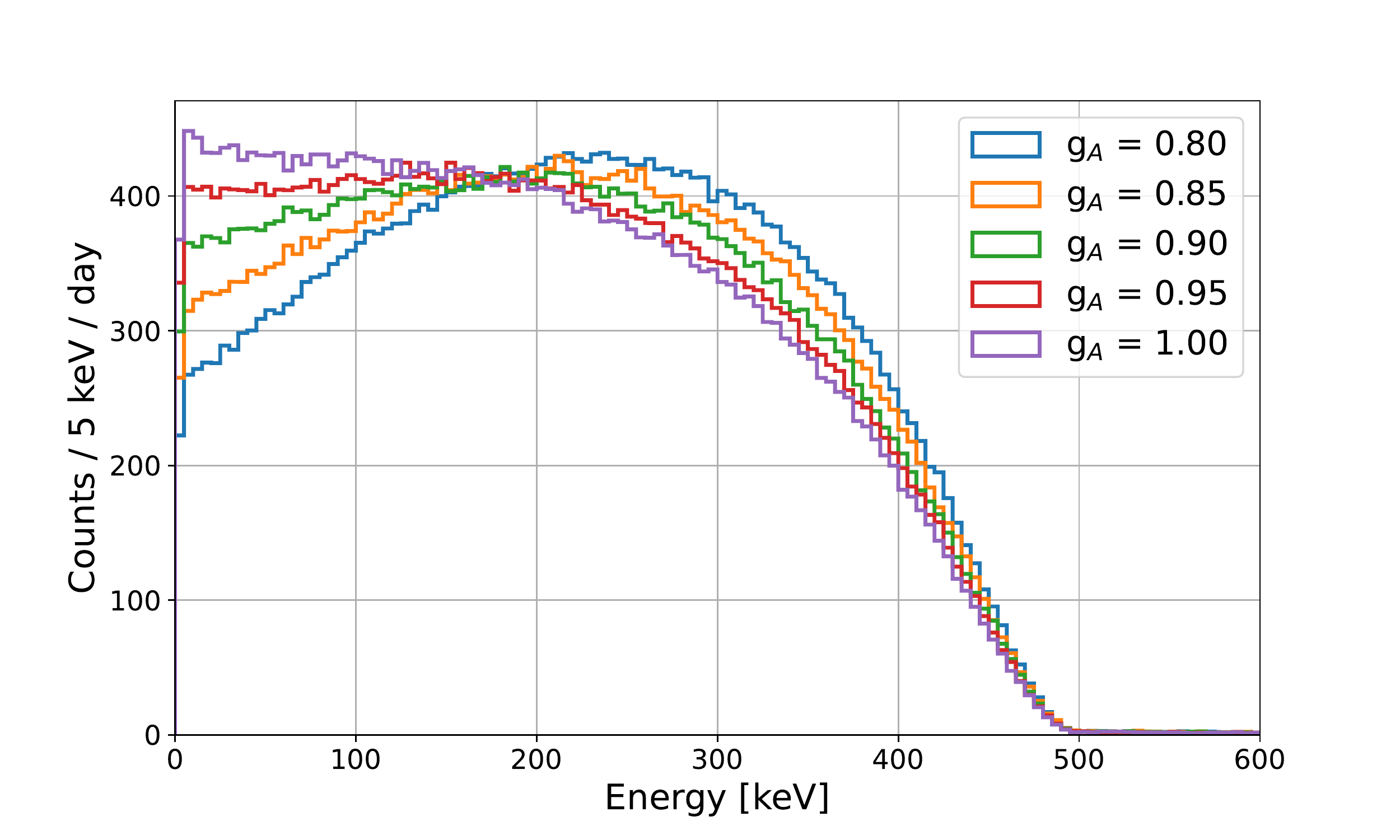}
	\caption{Simulated energy spectra of the $^{115}$In $\beta$-decay for different dimensions of the absorber (assuming s-NME $= 2.0$ and $g_{\rm A} = 0.9$, left). The larger is the crystal the lower is the difference between the template spectrum (black) and the simulated one. The simulation for the 7~mm side crystal is repeated for five different values of $g_{\rm A}$ around the chosen reference value (right). The simulations are processed introducing the detector performance reported in Tab.~\ref{tab:detectorPerformance}}
	\label{fig:differentGA}
\end{figure}

Another crucial point of this study is the evaluation of the detector response function on the $\beta$-decay spectrum, because we want to avoid any distortions introduced for example by the pile-up. This is important to understand if the thermal sensors we are taking into account, NTD and TES, are suitable for our purposes. First of all, we want to minimise the partial containment of the signal, to be sure that the detector response function does not deteriorate our sensitivity to different $g_{\rm A}$. In Fig.~\ref{fig:differentGA}, we show $^{115}$In spectra generated for different sizes of the absorber (left), and different axial coupling constants (right). In order to be conservative, we look at them by considering in the Monte Carlo simulations the typical parameters of the sensor with poorer performance, i.e. the NTD reported in Tab.~\ref{tab:detectorPerformance}. Even if the detector response function can be effectively evaluated by Monte Carlo simulations, as demonstrated in several works~\cite{Azzolini:2019nmi}, we can conclude that larger crystals are preferred in order to minimise the effects due to partial containment, thus reducing any possible systematic uncertainty due to the detector response function.

\begin{figure}[ht]
	\centering
	\includegraphics[width=0.48\textwidth]{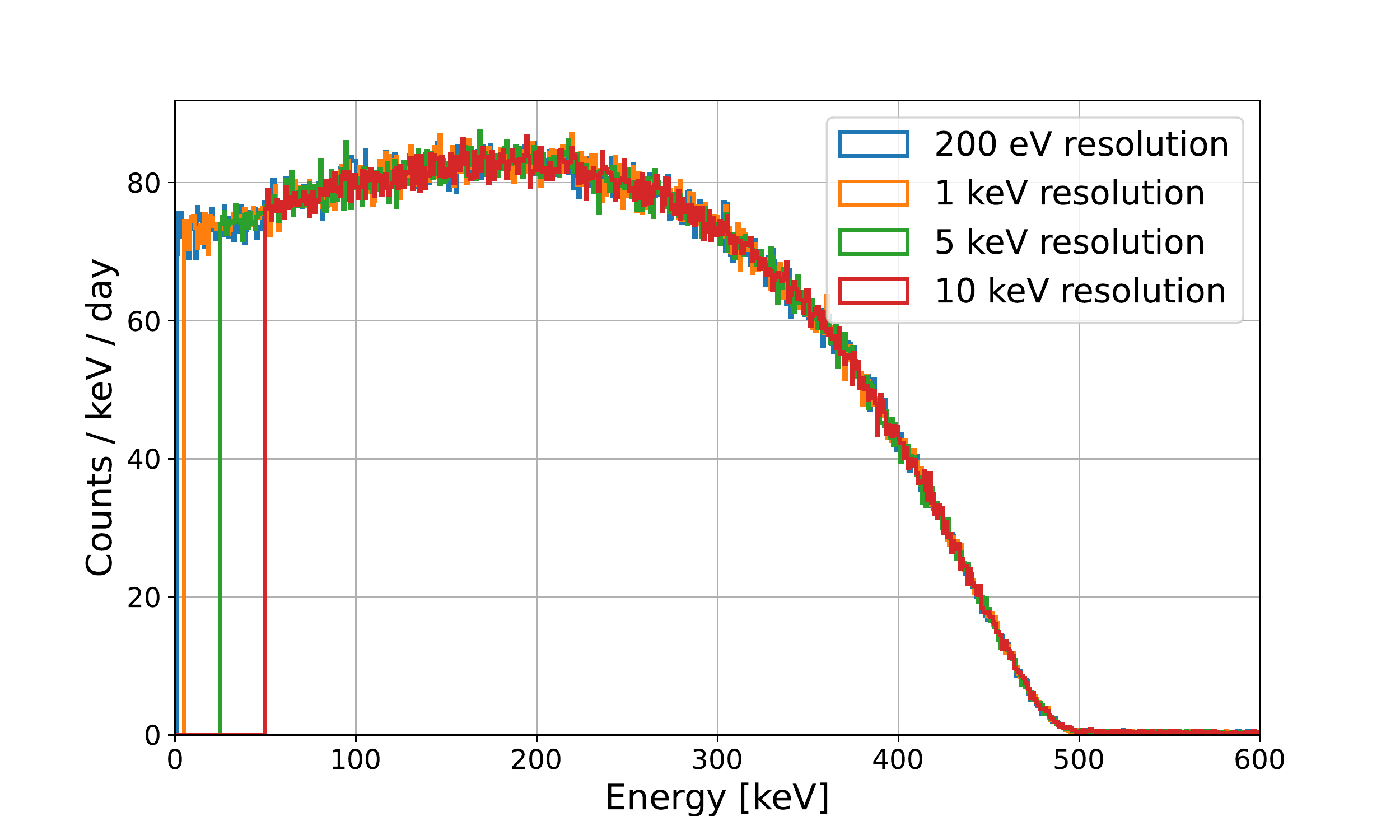}
	\caption{Effect of different energy resolution and energy thresholds on the $^{115}$In spectrum. Notice that the first parameter has no repercussions on the resulting shape}
	\label{fig:differentERes}
\end{figure}

Two other essential parameters to be considered are the energy resolution and the energy threshold. Typically the better the energy resolution is, the lower the detector energy threshold will be. Let us fix the theory input as before: s-NME $= 2.0$ and $g_{\rm A} = 0.90$, the time resolution (10 ms) and the acquisition window (1 s). We take into account 4 different constant energy resolutions: \SI{200}{eV}, \SI{1}{keV}, \SI{5}{keV} and \SI{10}{keV}.
As a rule of thumb, we assign an energy threshold equal to 5 times the energy resolution, so respectively \SI{1}{keV}, \SI{5}{keV}, \SI{25}{keV} and \SI{50}{keV}. The simulated spectra processed with these parameters are presented in Fig.~\ref{fig:differentERes}. A worsening in the energy resolution is not affecting the spectral shape, our signal being a continuum spectrum. Nevertheless, a better energy resolution helps to fix the different background sources in the spectrum reconstruction.
Conversely, an energy threshold larger than 50 keV cuts away a good fraction of the low energy region, where the spectral shape effects due to a different value of $g_{\rm A}$ are more relevant. Thus, both the thermal sensors considered in this paper are suitable for these studies as long as the energy threshold is lower than 50 keV.

\begin{figure}[ht]
	\centering
	\includegraphics[width=0.48\textwidth]{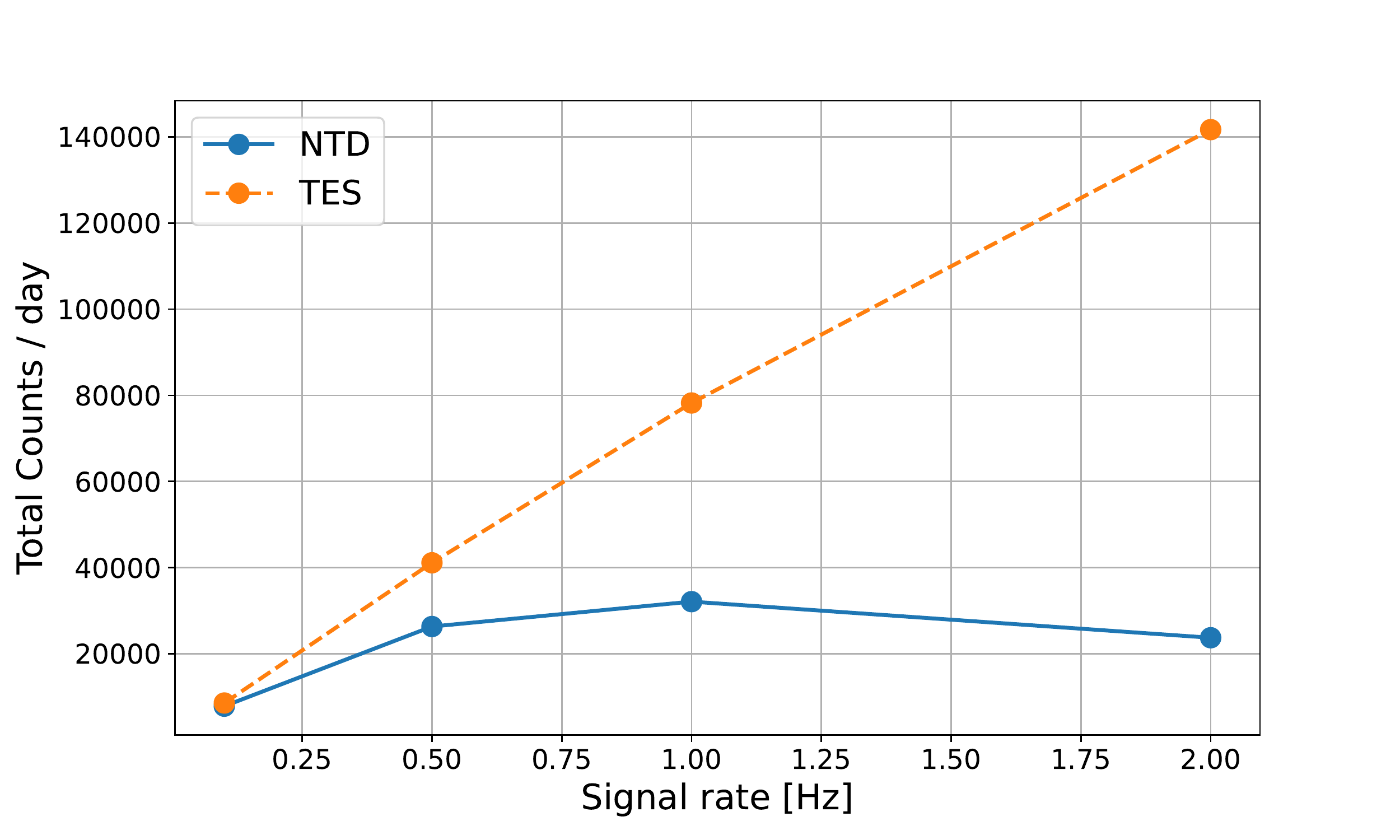}		\includegraphics[width=0.48\textwidth]{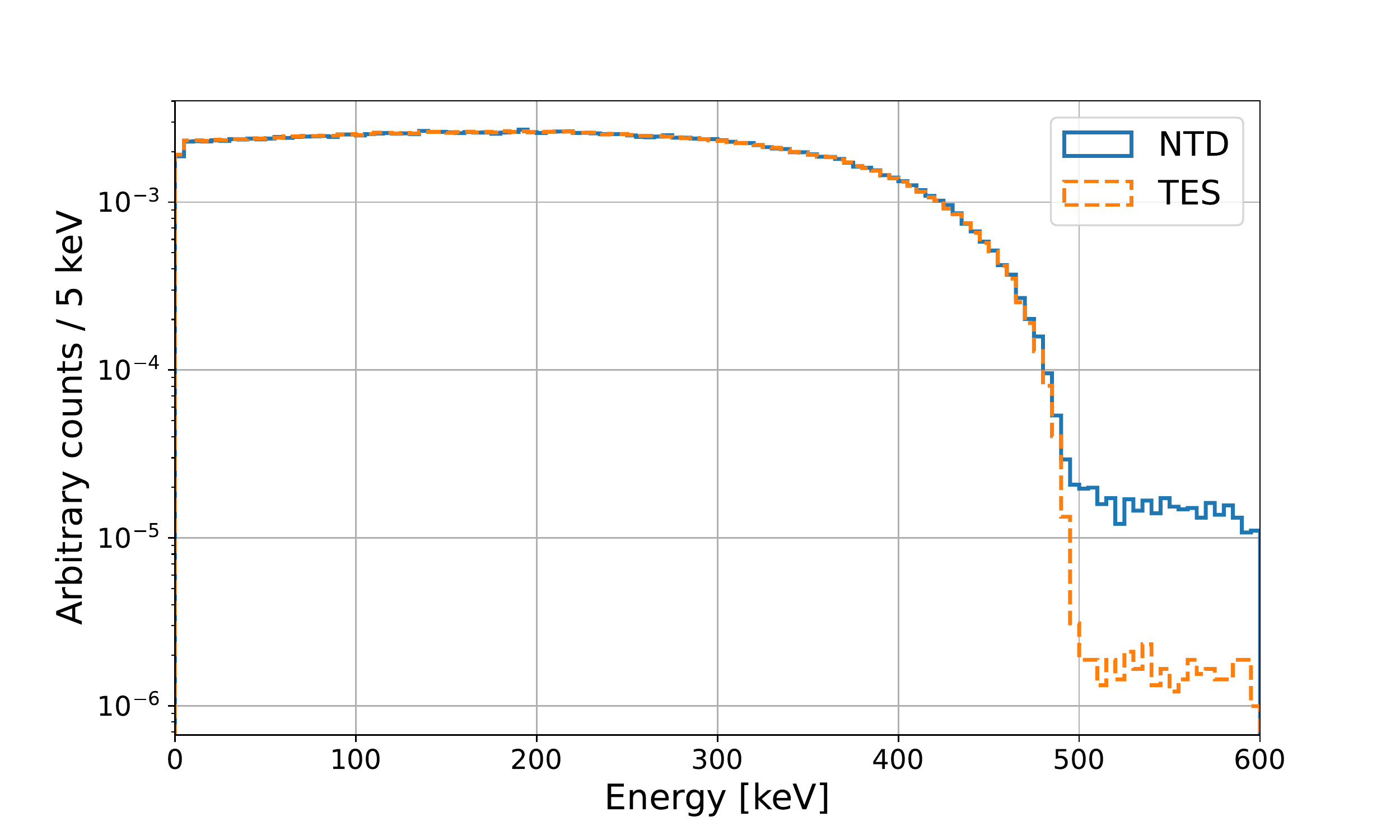}
	\caption{{\it Left.} Expected signal counts per day as a function of the source activity assuming time and energy resolutions of a NTD (blue solid line) and a TES (orange dashed line). The divergence from the linear trend grows with the expected signal rate due the presence of multiple signals in the same acquisition time window.
    {\it Right.} Energy spectra of the $^{115}$In $\beta$-decay assuming time and energy resolutions of a NTD (blue solid line) and a TES (orange dashed line). As expected, the piled up events populate the energy spectrum above the $^{115}$In Q-value in the NTD case, while the pileup is practically negligible for the TES}
	\label{fig:Integrals+NTDvsTES}
\end{figure}

Lastly, we study the effects of different timing parameters, i.e. the time resolution and the acquisition window length. A good time resolution allows us to distinguish pulses close in time that would be integrated together otherwise. Moreover, faster pulses ensure shorter acquisition intervals, reducing the statistical loss due to multiple pulses in the same window, usually discarded during the data analysis.
TES operated with macro-calorimeters of the size mentioned above, usually have rise times in the order of \SI{1}{ms}. We decide cautiously to consider a time resolution equal to the rise time itself since the resolution heavily depends on the algorithm used to tag overlapped pulses, even if it can be reduced up to 10 times with sophisticated analyses~\cite{Nucciotti_2018}.
With a signal rate of $\gtrsim$ \SI{1}{Hz}, the presence of multiple pulses within a signal time window reduces the actual signal rate for the NTD configuration, while TES are practically unaffected, as reported in Fig.~\ref{fig:Integrals+NTDvsTES} (left). This could represent an issue when the size of natural crystals is too large or the source activity in doped crystals is too high. As shown in Fig.~\ref{fig:Integrals+NTDvsTES} (right), the different time resolutions considered do not introduce shape distortions to the energy spectrum, with the exception of the region near the decay Q-value, where this effect can be effectively modelled anyway. 

To conclude, both TES and NTD are suitable for the ACCESS goals with the caveat that energy threshold and statistics losses due to poor time resolution may represent limiting factors for the second class of thermal sensors. In the case of InI, the optimal crystal size determines an activity way lower than 1 Hz, so the NTD sensor is still a good option; conversely for doped crystals we can increase the signal-to-background ratio optimising the doping level, which requires a TES if the signal rate exceeds the 1 Hz limit. 

\section{Preliminary measurements}\label{Sec:Measurements}
\begin{figure}[ht!]
    \centering
    \includegraphics[width=0.33\textwidth]{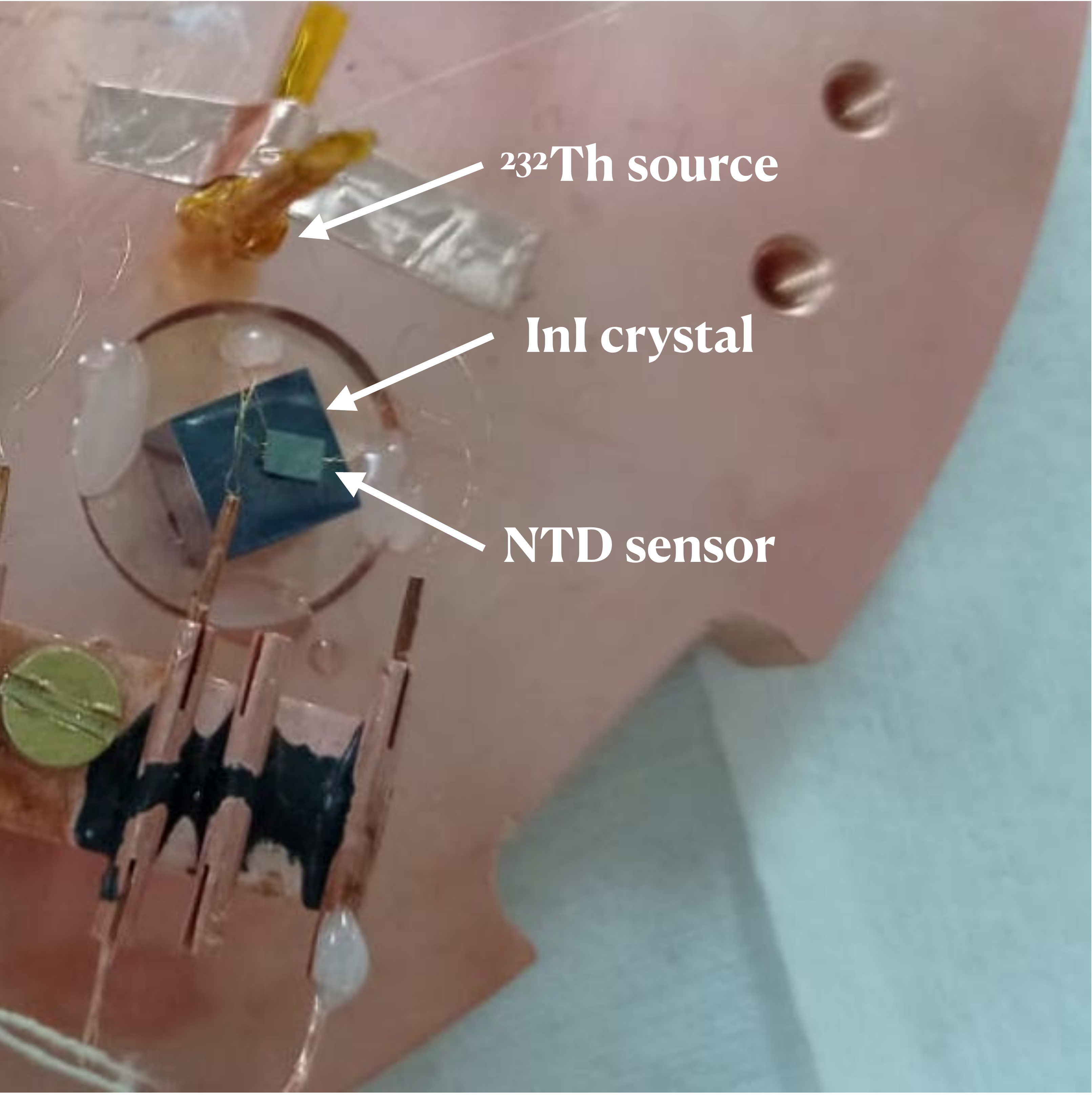}
    \caption{Experimental setup used to measure the InI crystal as cryogenic calorimeter at LNGS. The detector is resting on the copper plate facing a $^{232}$Th source}
    \label{fig:in_setup}
\end{figure}

\begin{figure}[ht]
    \centering
    \includegraphics[width=0.48\textwidth]{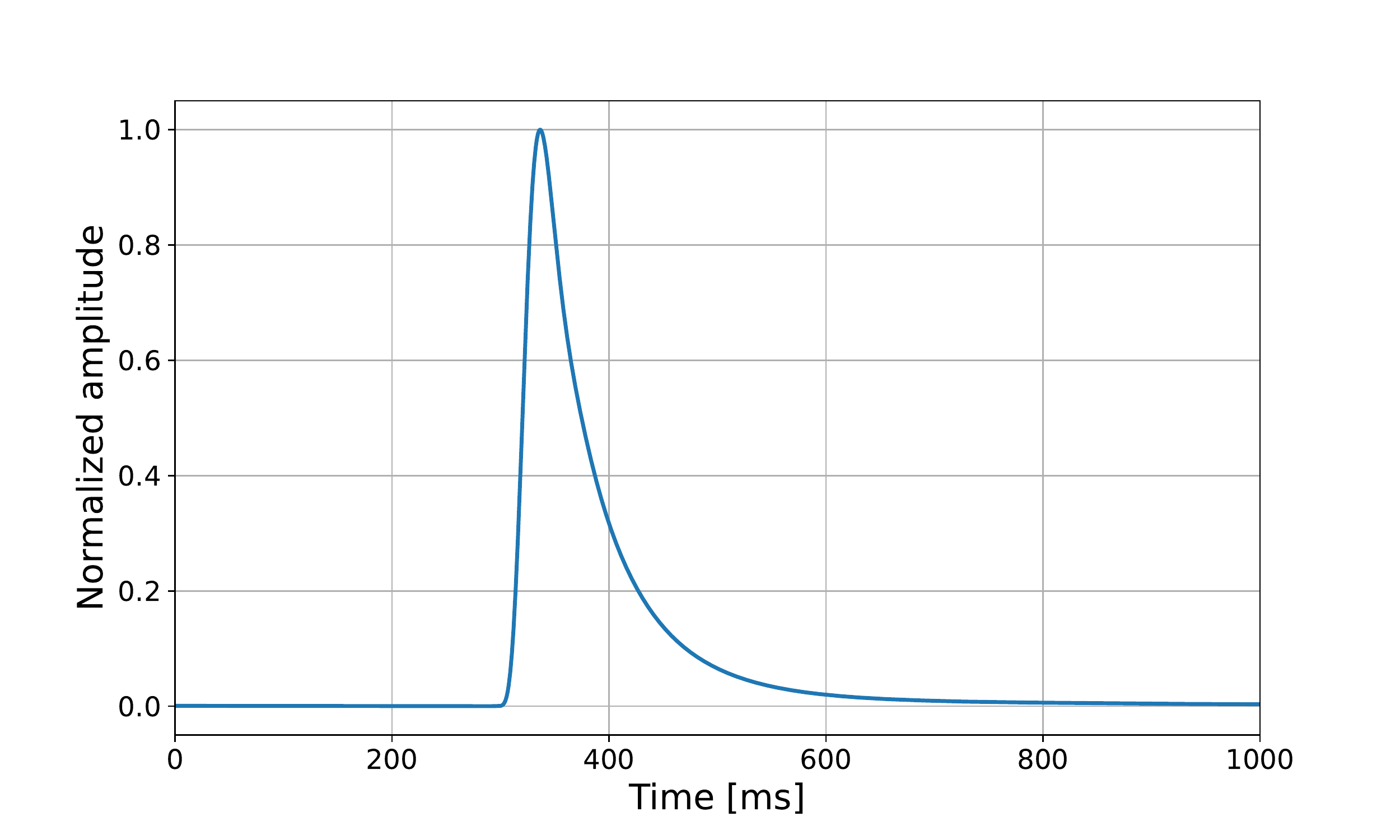}
    \includegraphics[width=0.48\textwidth]{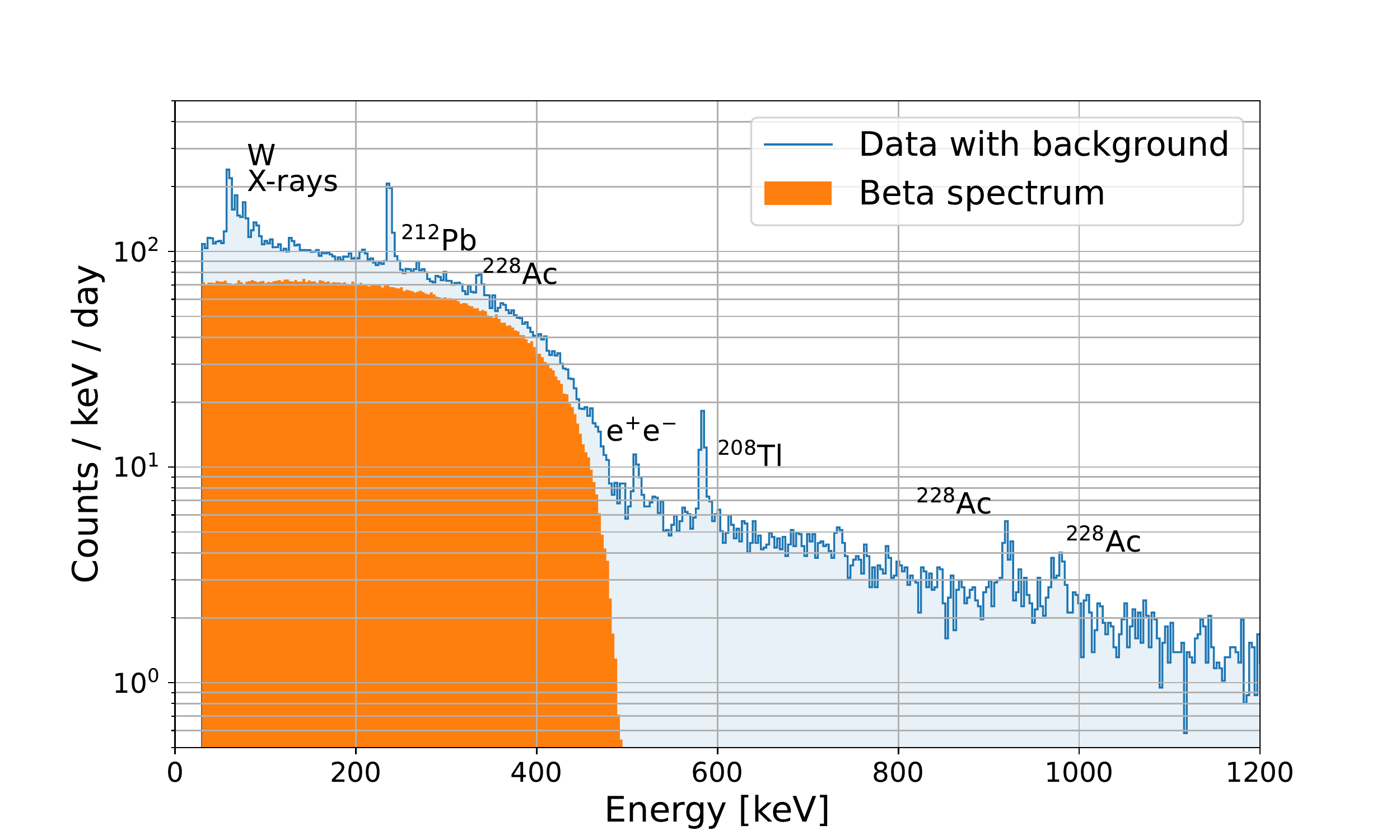}
    \caption{{\it Left.} Template signal obtained by averaging hundreds of InI thermal pulses using a 1000 ms acquisition window. 
    {\it Right.} Energy spectrum obtained with the InI crystal in 300 hours of data taking together with the $^{115}$In $\beta$-decay component (orange). The prominent $\gamma$-rays produced by a near $^{232}$Th source are labelled}
    \label{fig:Spectrum+Pulse}
\end{figure}

Following the outcomes of the study presented in Sec.~\ref{Sec:Study}, we tested an indium iodide (InI) crystal of dimensions 7$\times$7$\times$7 mm$^3$ and mass 1.9 g. We equipped the InI crystal with an NTD thermistor (3$\times$3$\times$1 mm$^3$), whose signal is readout with 25 $\mu$m gold wires. The detector was installed on a copper plate directly attached to the mixing chamber of the CUPID R$\&$D cryostat and operated at 16 mK. We also added a $^{232}$Th calibration source to the experimental setup, which is shown in Fig.~\ref{fig:in_setup}.\\
The electronic readout and the data acquisition systems used in this measurement are described in Refs.~\cite{Arnaboldi:2017aek,DiDomizio:2018ldc}. We saved on disk the continuum data stream, and every time the derivative trigger fired, a waveform of 1000 ms was recorded for InI. The collected data were then filtered with the optimum filter technique~\cite{Gatti:1986cw}, following the standard steps established and tested by CUORE-0~\cite{CUORE:2016ons}, CUORE~\cite{CUORE:2021mvw,CUORE:2020bok}, CUPID-0~\cite{Azzolini:2019yib,CUPID:2022puj}, and CUPID-Mo~\cite{Armengaud:2019rll,Augier:2022znx} collaborations.
In this configuration, we acquired over 300 h of physics data with a calibration source. The template pulse of the detector, obtained by averaging hundreds of signals, is presented in Fig.~\ref{fig:Spectrum+Pulse} (left).
We calibrate InI using the prominent $\gamma$-ray peaks in the energy spectrum as presented in Fig.~\ref{fig:in_setup}. Overall, the detector shows very good performance, achieving an energy threshold of 17 keV, and an energy resolution of 3.1 keV FWHM at 60 keV. Moreover, the dominant contribution in the spectrum is the $\beta$-decay of $^{115}$In, as presented in Fig.~\ref{fig:Spectrum+Pulse} (right). 

For the next steps of the project, we already established a collaboration to grow $^{99}$Tc-doped crystals with a company which produced for ACCESS two TeO$_2:^{99}$Tc crystals of dimensions 40$\times$35$\times$22 mm$^3$ and 35$\times$35$\times$22 mm$^3$, exploiting a $^{99}$Tc-doped TeO$_2$ powder. At LNGS, we performed ICP-MS analysis of doped powder, filtered powder ready for the crystal growth and final crystals, measuring a $^{99}$Tc concentration of $(8 \pm 2)$ ppb, $(2.2 \pm 0.2)$ ppb, and $< 0.1$ ppb, respectively. This preliminary material screening demonstrated that both powder filtering and crystal growth {\it purify} the crystal from $^{99}$Tc. We also operated the two TeO$_2:^{99}$Tc crystals as cryogenic calorimeters~\cite{Helis@MEDEX}, obtaining from the data analysis an upper limit on concentration of $^{99}$Tc of $0.01$ ppb, well below the target value. Further tests and investigations are currently ongoing to understand the origin of the $^{99}$Tc concentration reduction during the crystal production.

\section{Conclusions}\label{Sec:Conclusions}
In this paper, we presented for the first time the ACCESS project, whose main goal is to develop cryogenic calorimeters to asses the spectral shape of forbidden $\beta$-decay. As first step, we proposed to measure the $^{115}$In $\beta$-decay exploiting an InI natural crystal. The studies reported in Sec.~\ref{Sec:Study} on InI demonstrate that a 7-mm side cubic crystal is an excellent compromise to optimise simultaneously the signal yield and the detector response function, avoiding efficiency loss induced by a high counting rate. Moreover, we proved that the time resolution of an NTD sensor ($\sim$10 ms) is suitable for this kind of crystal. On these bases, we tested an InI-based cryogenic calorimeter obtaining an excellent threshold of 17 keV, and an energy resolution of 3.1 keV FWHM at 60 keV.
This achieved performance matches the ACCESS requirements and a further measurement of the InI detector with higher statistics is on the schedule.

\section*{Acknowledgements}
This project has received funding from the European Union’s Horizon 2020 research and innovation program under the Marie Skłodowska–Curie grant agreement N.~101029688. This work was supported by the Academy of Finland, Grant Nos.~314733, 320062 and 345869. We thank the CUPID collaboration for sharing their cryogenic infrastructure, M. Guetti for the assistance in the cryogenic operations, M. Perego for his invaluable help in many tasks, the mechanical workshop of LNGS. This work makes use of the DIANA data analysis and APOLLO data acquisition software which has been developed by the CUORICINO, CUORE, LUCIFER and CUPID-0 collaborations.
\begin{figure}[h!]
    \centering
    \includegraphics[width=0.2\textwidth]{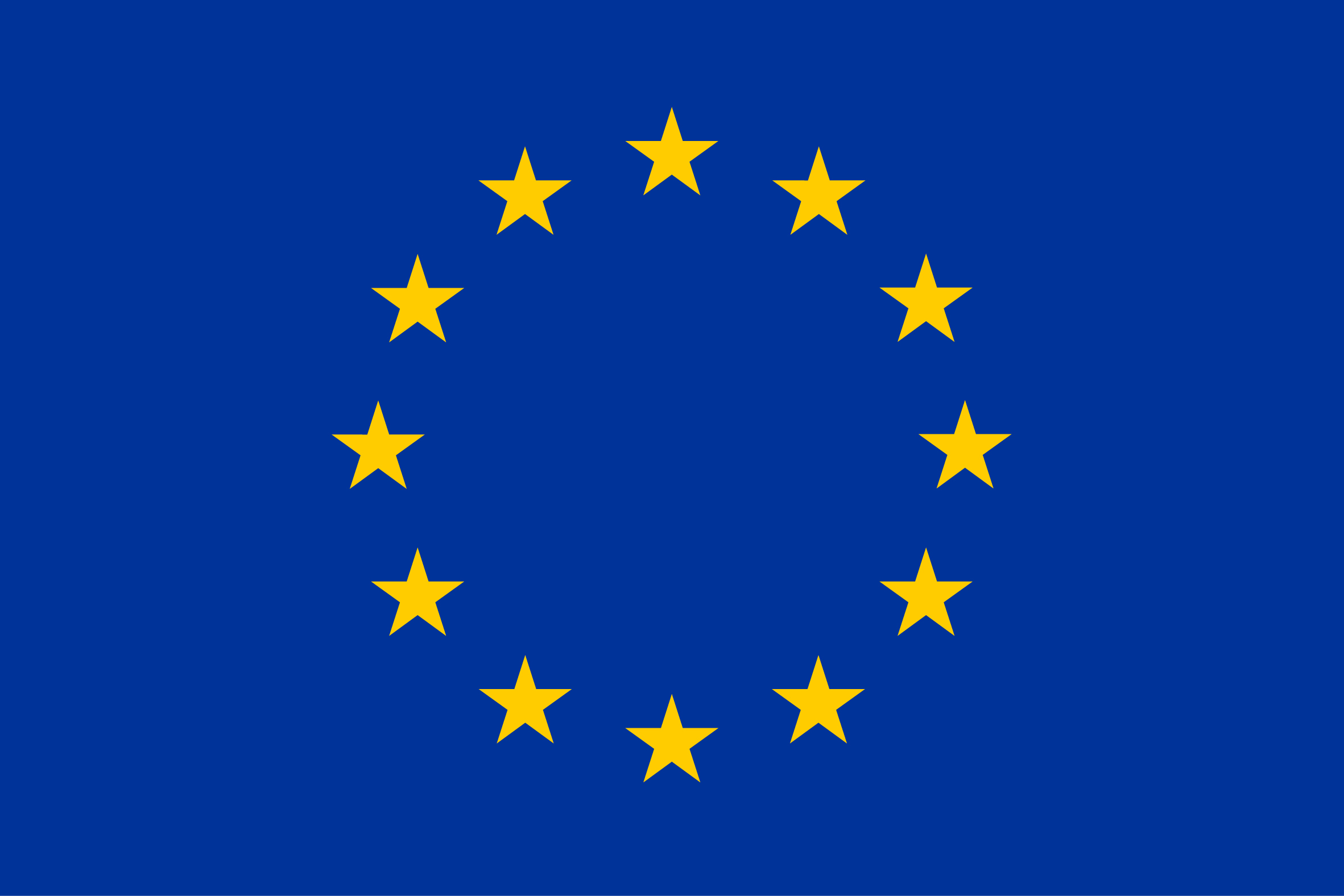}
\end{figure}

\bibliography{Bibliography}
\bibliographystyle{spphys}
%
%
%

\end{document}